\documentclass[fleqn,usenatbib]{mnras}

\usepackage{graphicx,natbib,times} 
\usepackage[T1]{fontenc}
\DeclareRobustCommand{\VAN}[3]{#2}
\let\VANthebibliography\thebibliography
\def\thebibliography{\DeclareRobustCommand{\VAN}[3]{##3}\VANthebibliography}
\usepackage{amsmath}	

\newcommand{\microhz}{$\mu$Hz}

\title[BiSON farside helioseismology]{Farside helioseismology with Sun-as-a-star data: the solar cycle as seen with 7-day-long BiSON timeseries}
\author[R. Howe et al.]{
R. Howe,$^{1}$\thanks{E-mail: r.howe@bham.ac.uk (RH)}
W.~J.~Chaplin,$^{1}$
Y.~P.~Elsworth,$^{1}$
S.~J.~Hale,$^{1}$
E.~Hatt,$^{1}$
and M.~B.~Nielsen$^{1}$
\\
$^{1}$School of Physics and Astronomy, University of Birmingham, Edgbaston, Birmingham B15 2TT, UK} 

\date{Accepted XXX. Received YYY; in original form ZZZ}

\pubyear{2025}

\begin{document}
\label{firstpage}
\pagerange{\pageref{firstpage}--\pageref{lastpage}}
\maketitle

\begin{abstract}
We present results from fitting $p$-mode spectra derived from 7-d segments of Sun-as-a-star helioseismic observations from the Birmingham Solar Oscillations Network covering 32 yr. The results show a clear dependence of the mode frequencies on solar activity, and the frequency dependence of the sensitivity to activity can also be seen. Because we use data segments that cover less than half of a solar rotation, we are able to test for the effect of activity on the solar far side. By fitting with a model that takes into account activity on the far side of the Sun, we show that the frequency shifts are sensitive to activity from the whole Sun, not just the side facing the observer. Our results suggest that there is potential to investigate activity-related asteroseismic frequency shifts in solar-like oscillators using short time series of observations.

\end{abstract}
\section{Introduction}

The frequencies of solar acoustic modes vary with magnetic activity, with an average shift of a few parts in ten thousand between solar maximum and solar minimum, or about one part in a million per unit change in the sunspot number. This has been known since the early days of systematic observations of the oscillations; for example it was noticed by \cite{1985Natur.318..449W} and \cite{1991Natur.351...42W} in observations using solar irradiance data, and also reported by \cite{1989A&A...224..253P} and 
\cite{1990Natur.345..322E} in integrated-Sun observations and
by \citet{1990Natur.345..779L} for resolved-Sun observations from the Big Bear Solar Observatory. For global modes, the central frequencies of rotationally-split multiplets are well correlated with global activity indices, but the variation also exists at smaller spatial scales. For example, \cite{1988ApJ...331L.131K} found that the even-order coefficients of the expansion of the frequencies of a rotationally split multiplet, which describe the asphericity of the cavity in which the modes propagate, correlated with the aspherical temperature variations at the solar limb. \cite{2002ApJ...580.1172H} showed that the frequency shifts of individual multiplet components from Global Oscillations Network Group \citep[GONG, ][]{1996Sci...272.1284H} observations of medium-degree global modes correlate well with the latitudinal distribution of unsigned surface magnetic flux, and they were able to use an inversion technique to map the frequency shifts back to the magnetic ``butterfly diagram.''  At even smaller spatial and temporal scales, \cite{2000SoPh..192..363H} showed that the frequencies of high-degree modes measured in 15-degree patches of the solar surface using the ring-diagram \citep{1988ApJ...333..996H} technique of local helioseismology were increased in the locations of active regions. There is some evidence that the sensitivity of the mode frequencies to activity can vary between different solar cycles \citep[e.g.,][and references therein]{2018MNRAS.480L..79H} 
or on different timescales \citep[e.g.,][]{2007SoPh..243..105T}.

Frequency shifts are normally compared with activity proxies, such as the sunspot number,  10.7\,cm radio flux, or longitudinal magnetic field strength, that necessarily measure activity on the side of the Sun facing the observer. Because the acoustic modes we measure at low to medium degree are global standing waves, unlike the high-degree modes studied by \cite{2000SoPh..192..363H}, their frequencies should be sensitive to activity at all solar longitudes. Indeed, \citet{2001ApJ...560L.189B} developed a technique for imaging active regions on the solar far side using helioseismic holography applied to near-side observations from imaging instruments such as GONG and the Michelson Doppler Imager \citep[MDI, ][]{1995SoPh..162..129S}. More recently the same technique has been applied to the data from the Helioseismic and Magnetic Imager \citep[HMI, ][]{2012SoPh..275..229S}. Such images provide a useful tool for space weather forecasting.

Typically, global helioseismic analyses use Fourier spectra from observations covering multiple solar rotations, which makes the farside effects difficult to unravel. The aim of this work is to see whether it is practical to detect the effects of activity on the far side of the Sun by measuring the changing frequencies of low-degree global modes from observations spanning less than a solar rotation period. We use data from the Birmingham Solar Oscillations Network (BiSON), a six-site global network \citep{1996SoPh..168....1C,2016SoPh..291....1H} making integrated-light Doppler velocity observations of the Sun in the Potassium D1 line. The full network was deployed in 1992, and since then the duty cycle of the observations has been around 75 per cent on average. If we consider only the data from 1992 onward, we have approximately 32 years of observations covering nearly three solar cycles.

\section{Data and Analysis}

\subsection{``Near-side'' activity proxy}

For a solar-activity proxy we used the 10.7\,cm radio flux \citep[RF,][]{2013SpWea..11..394T} from 
\url{https://www.spaceweather.gc.ca/forecast-prevision/solar-solaire/solarflux/sx-5-flux-en.php}. This proxy is known to correlate well with global helioseismic frequencies on timescales of months \citep[e.g..][]{1993ApJ...411L..45B, 2001MNRAS.322...22C, 2012ApJ...758...43B,2017MNRAS.470.1935H}; it also has the advantage of being available on a daily cadence covering the entire span of the BiSON observations. 
To form the ``near-side'' proxy for a given observing period we take a mean of the daily RF values over the period, weighted by the BiSON duty cycle for each day.

\subsection{A ``far-side'' activity proxy}

\begin{figure*}
\centering
\includegraphics[width=0.9\linewidth]{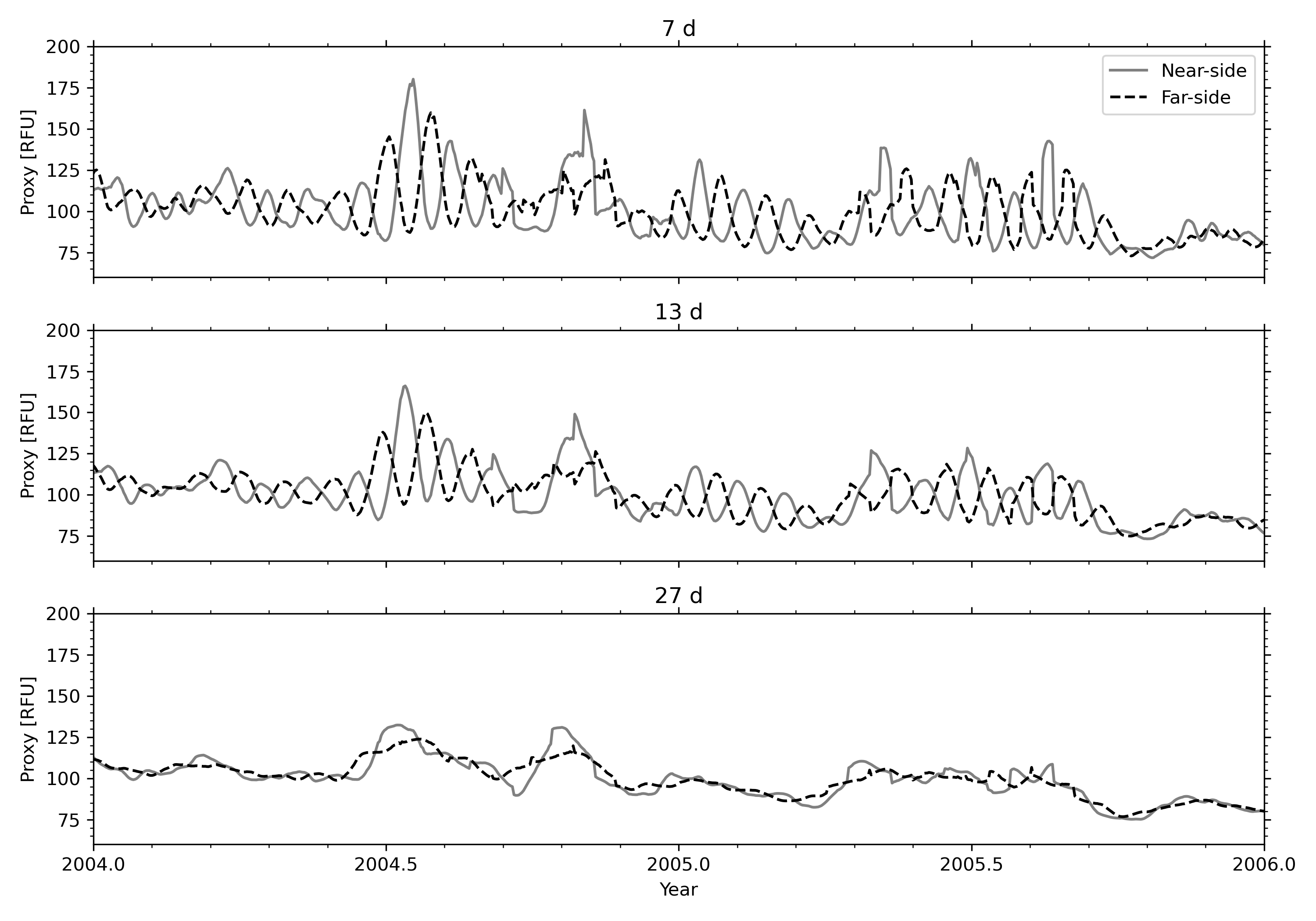}
\caption{Near- and far-side RF proxies for observing periods of 7, 13, and 27 days, plotted over a two-year period in the declining phase of Solar Cycle 23.}
\label{fig:proxy}
\end{figure*}

When we analyse the data in segments shorter than a solar rotation, our activity proxy also covers less than a rotation and hence does not encompass all of the activity that is happening on the Sun, but only that on the side facing the Earth during the observation period. When activity is concentrated in a particular range of longitudes, this results in the overall activity being underestimated when the active longitudes are out of view and overestimated when they face the observer. Global modes, however, should be sensitive to the full range of longitudes. As a crude proxy for the far-side activity, we use the average of the ``near-side" proxy over periods 13\,d earlier and 13\,d later than the observing period for the oscillations. For example, for a 7-day observing period starting on day 0, we would average the RF index over days -13 to -7 and days 13 to 19, but weighted by the BiSON duty cycle for days 0 to 6. This should account for activity that has been seen on the Earth-side and rotated out of view as well as activity that has emerged on the far side and will rotate into view in the future, but it does not include any activity that emerges and decays out of view and never rotates onto the visible side of the Sun.  Figure~\ref{fig:proxy} shows this far-side proxy calculated for different observation lengths and plotted on the same axes as the normal near-side RF value. For short spectra, the two proxies are clearly distinct on timescales of a few months and even vary in antiphase at times, reflecting epochs when activity was concentrated on one side of the Sun. As the observation length approaches a full rotation the two proxies become almost indistinguishable.

\subsection{BiSON data}

The helioseismic data for this project consist of the BiSON residual velocity time series for the period 1992\,--\,2023 covering the declining phase of Solar Cycle 22, the whole of Cycles 23 and 24, and the rising phase of Cycle 25. The time series was prepared as described by \citet{doi:10.1093/mnras/stu803}. We analysed segments of varying lengths, from 27\,d down to 7\,d. The 27-d interval was chosen to correspond to the solar rotation period.
For each length $N$, 
a series of Fourier power spectra was constructed from $N$-d segments of the time series with start times at 1-d intervals, excluding periods where the duty cycle was less than 40 per cent. 
For comparison, we also considered frequency shifts from 365-day time series, similar to those analysed by \citet{2015MNRAS.454.4120H}.

The mode frequencies were estimated by fitting each spectrum using a slightly modified version of the Bayesian fitting algorithm described by \cite{10.1093mnrasstad2753}. This algorithm was originally designed for deriving rotational splittings from very long time series. In the short-interval spectra used for the present work the rotational splitting is not resolved, and we are interested only in the mean shift of the multiplet frequency. The splitting value was therefore fixed (by using very narrow Gaussian priors) at 400\,nHz, to give a faster and more stable fit.

\subsection{Analysis}

For each time series segment, frequencies were obtained for modes between $n=12$ and $n=25$, $0 \leq l \leq 3$.
We then found the frequency shift of each mode in each set relative to the mean for that mode over the whole set of segments of the same length and averaged the shifts for the modes with $l\leq 2$ and $n\geq 19$, which have the clearest signature of variation with activity level. The $l=3$ modes were excluded from the average because they have lower signal to noise ratio than the other modes, and the lower-frequency modes because they have lower sensitivity to solar activity.

\section{Results}

\begin{figure*}
    \centering
       \includegraphics[width=0.9\linewidth]{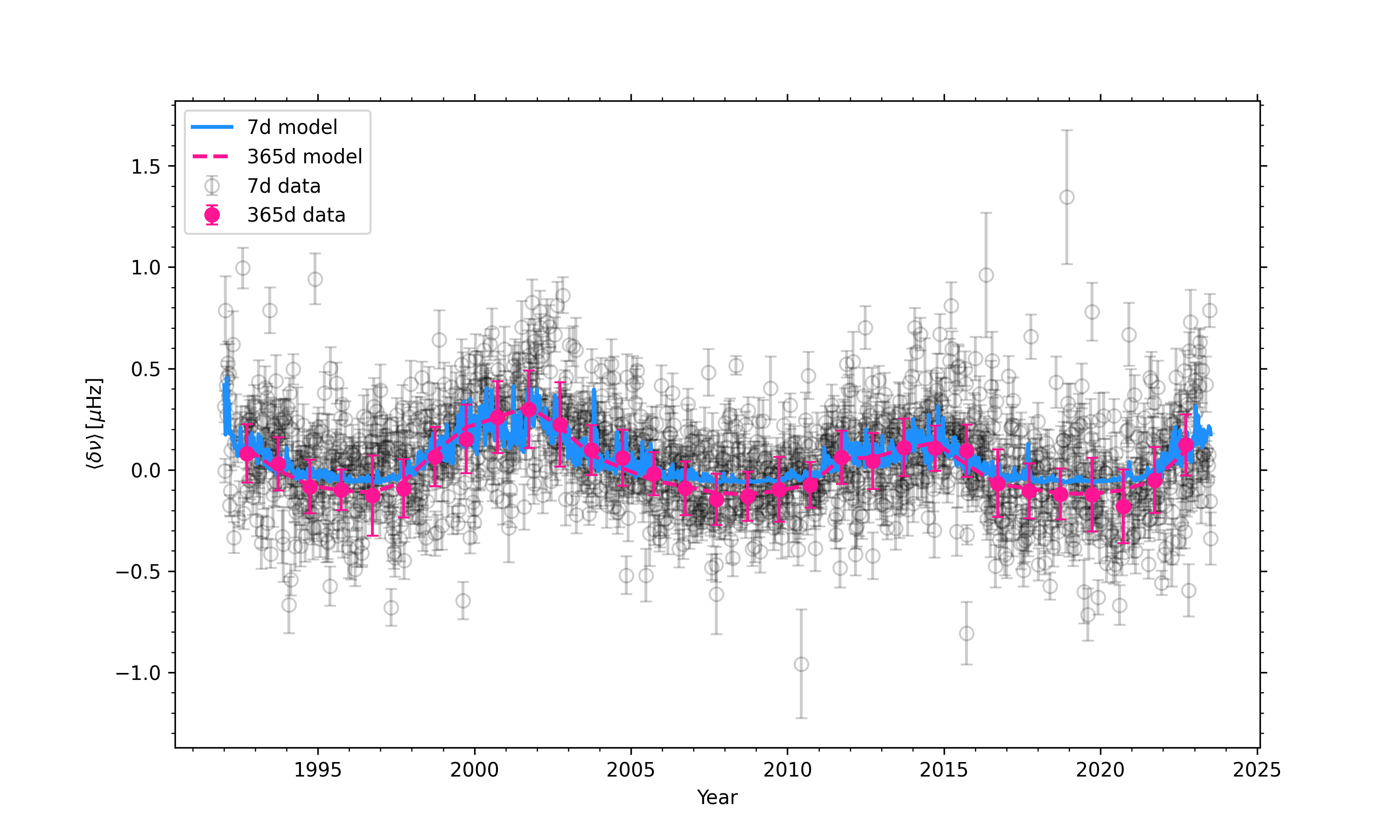}
    
    \caption{Mean frequency shift for modes with $n\geq 19, l \leq 2$ for non-overlapping 7-d BiSON spectra
    and 365-day spectra. Note that 10-$\sigma$ error bars are shown for the 365-day shifts, to make the points more visible. The curves show the best linear fit to the standard ``near-side'' RF index for each dataset.}
    \label{fig:fig3}
\end{figure*}

Figure~\ref{fig:fig3} shows the mean frequency shifts $\langle\delta\nu\rangle$ for $n \geq 19, l \leq 2$ as a function of time 
for 7-d and 365-d spectra. Also plotted is a curve representing a linear fit of the mean shifts to the equation 
\begin {equation}
\langle\delta\nu\rangle = S P_{\mathrm{near}} + c,
\label{eq:avshift}
\end{equation}
where $P_{\mathrm{near}}$ 
is the near-side RF index, $S$ is a factor representing the sensitivity of the shifts to the activity proxy, and $c$ is a constant.

The data set contains 1607 7-d spectra. The mean uncertainty of the average shift is around 0.1 \microhz, which is about 20 per cent of the overall cycle variation for modes in this frequency range.

\begin{figure*}
    \centering
    \includegraphics[width=0.49\linewidth]{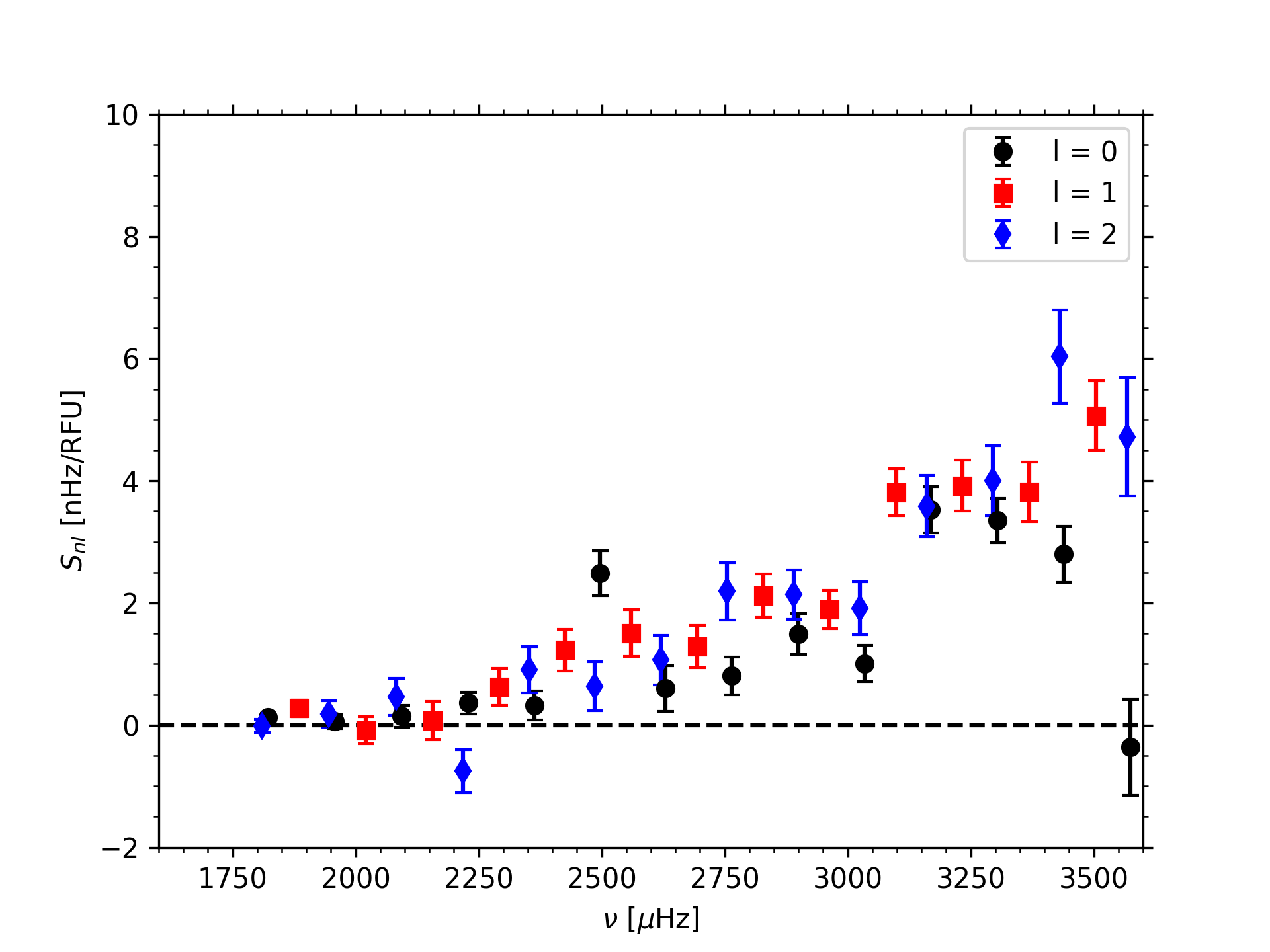}
    \includegraphics[width=0.49\linewidth]{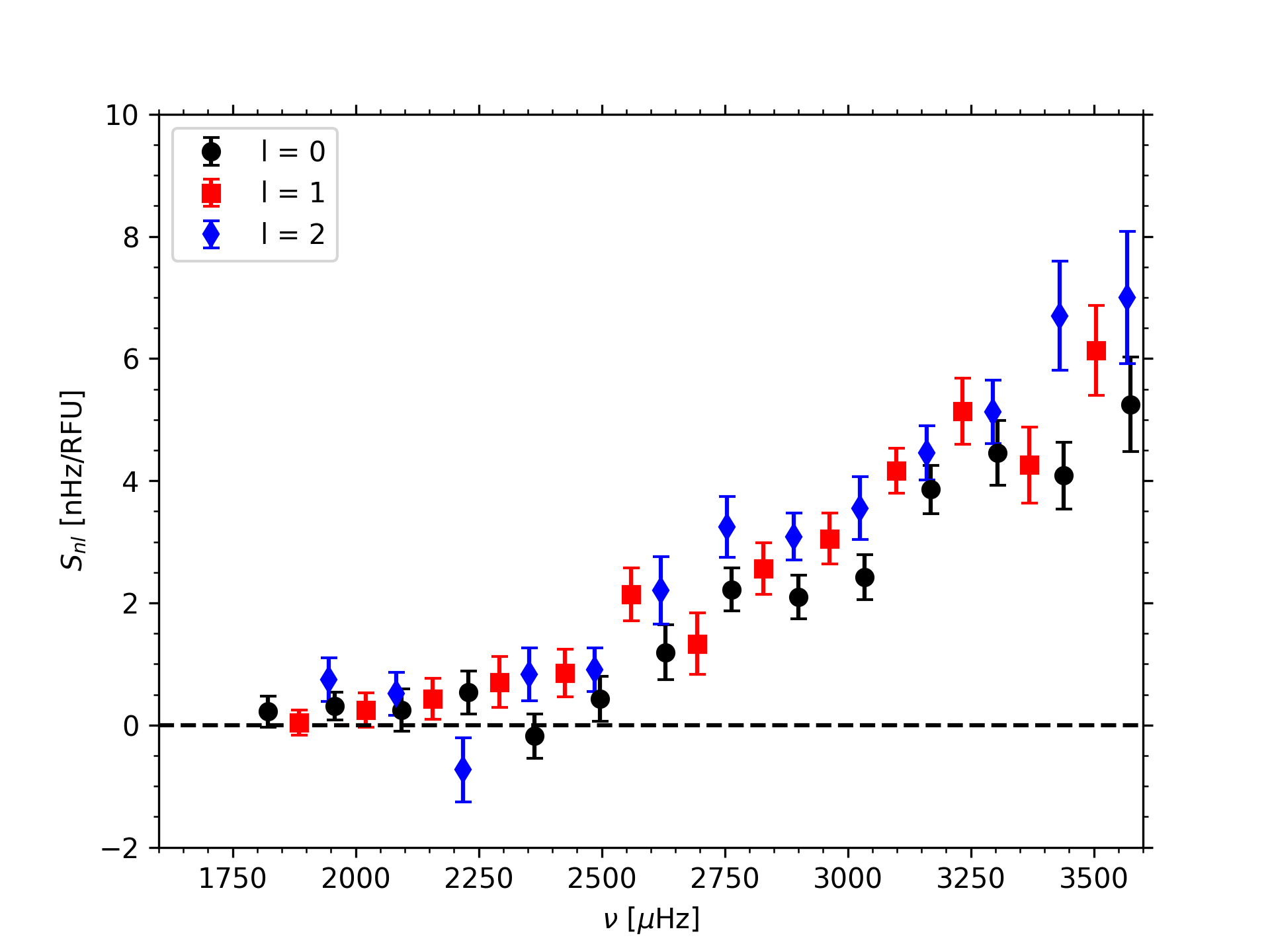}

    \caption{Mode-by-mode sensitivity of the BiSON frequency 
    shifts in 7-d (left) and 365-d (right) spectra to the ``near-side'' RF index, colour-coded by the degree $l$. The sensitivity values were calculated using only frequencies from spectra with no temporal overlap.
    } 
    \label{fig:fig4a}
\end{figure*}

We can analyse the shifts for individual modes by fitting the frequency for each mode to a linear function of the activity proxy, in a similar way to Equation~\ref{eq:avshift}. The frequency shift for a mode of radial order $n$ and degree $l$ is assumed to be given by
\begin{equation}
\delta\nu_{nl} = S_{nl} P_{\mathrm{near}} + c_{nl},
\label{eq:oneside}
\end{equation}
where $S_{nl}$ is a sensitivity term, $P_{\mathrm{near}}$ is the near-side activity index, and $c_{nl}$ is a constant corresponding to the 
value of the the frequency $\nu_{nl}$ when $P_{\mathrm{near}}$ is zero.
In Figure~\ref{fig:fig4a} we show $S_{nl}$
plotted as a function of frequency for 7-d and 365-d spectra. The sensitivity and its frequency variation are similar to those reported by \cite{2015MNRAS.454.4120H} for 365-d spectra, which gives us confidence that the fit is working reasonably well. Even for the 7-d spectra, the frequency variation of the sensitivity is clear.

\begin{figure*}    
\includegraphics[width=0.9\linewidth]{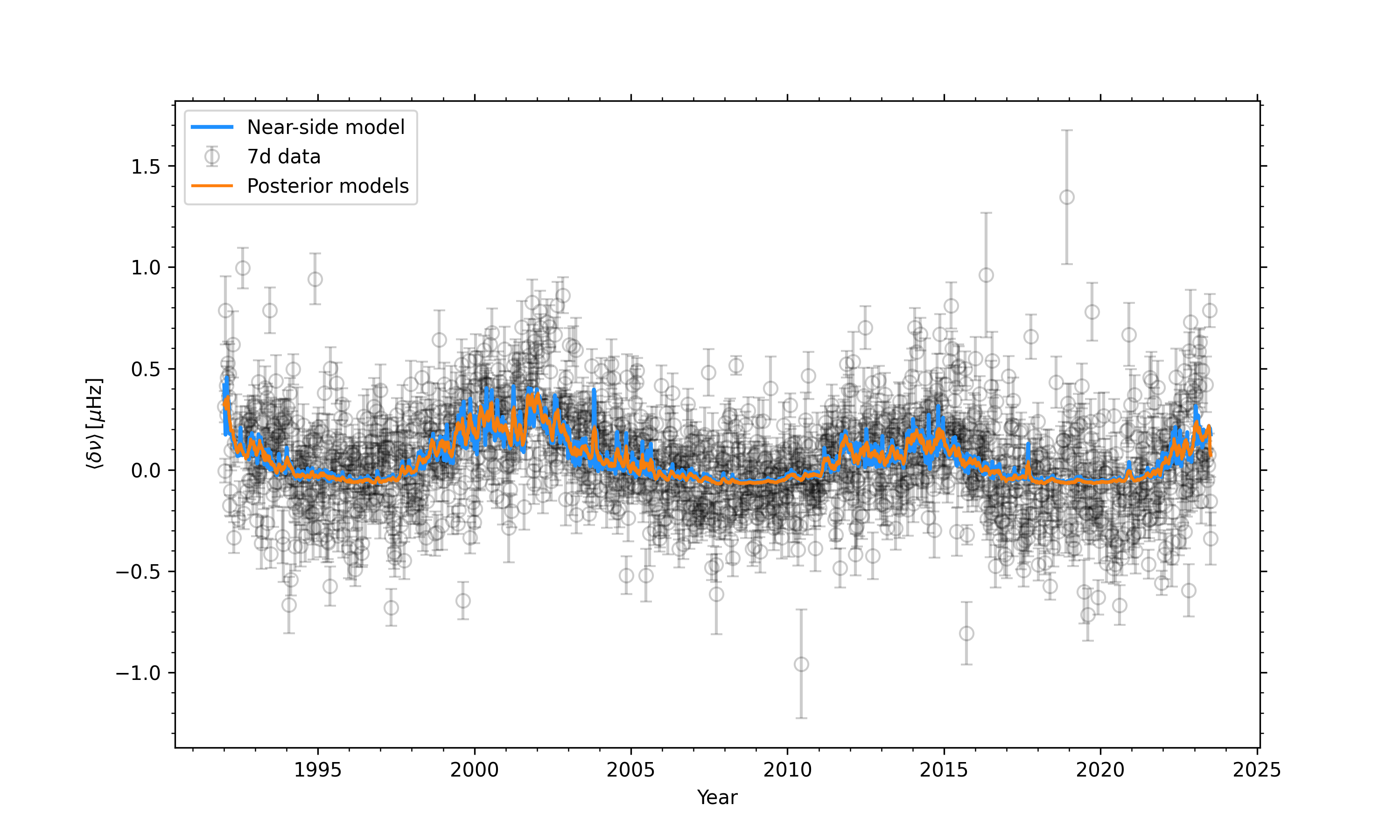}
\caption{Mean frequency shifts from non-overlapping 7-d BiSON spectra. The posterior models for the fit to Eq.~\ref{eq:twosides} are shown in orange and the linear fit to the near-side RF proxy is shown in blue.}
\label{fig:fig5a}
\end{figure*}

The formulation in Eq.~\ref{eq:oneside} neglects activity on the far side of the Sun. 
To test for the effect of activity that is not accounted for in the near-side index on the frequencies from 7-d spectra, we fit a model in which the average frequency shift $\langle\delta\nu\rangle$ is assumed to be linearly dependent on a combination of the near-side and far-side RF indices:
\begin{equation}
    \langle\delta\nu\rangle= S[(1-\beta) P_{\mathrm{near}} + \beta P_{\mathrm{far}}] + c
\label{eq:twosides}
\end{equation}
where $S$ is a sensitivity term for the mean shift, $\beta$ is the parameter controlling the relative contributions of near-side ($P_{\mathrm{near}}$) and far-side ($P_{\mathrm{far}}$) activity, and $c$ is a constant. This fit was carried out using the emcee package \citep{2013PASP..125..306F}. Figure~\ref{fig:fig5a} shows the fitted model.

\begin{figure}
\includegraphics[width=\linewidth]{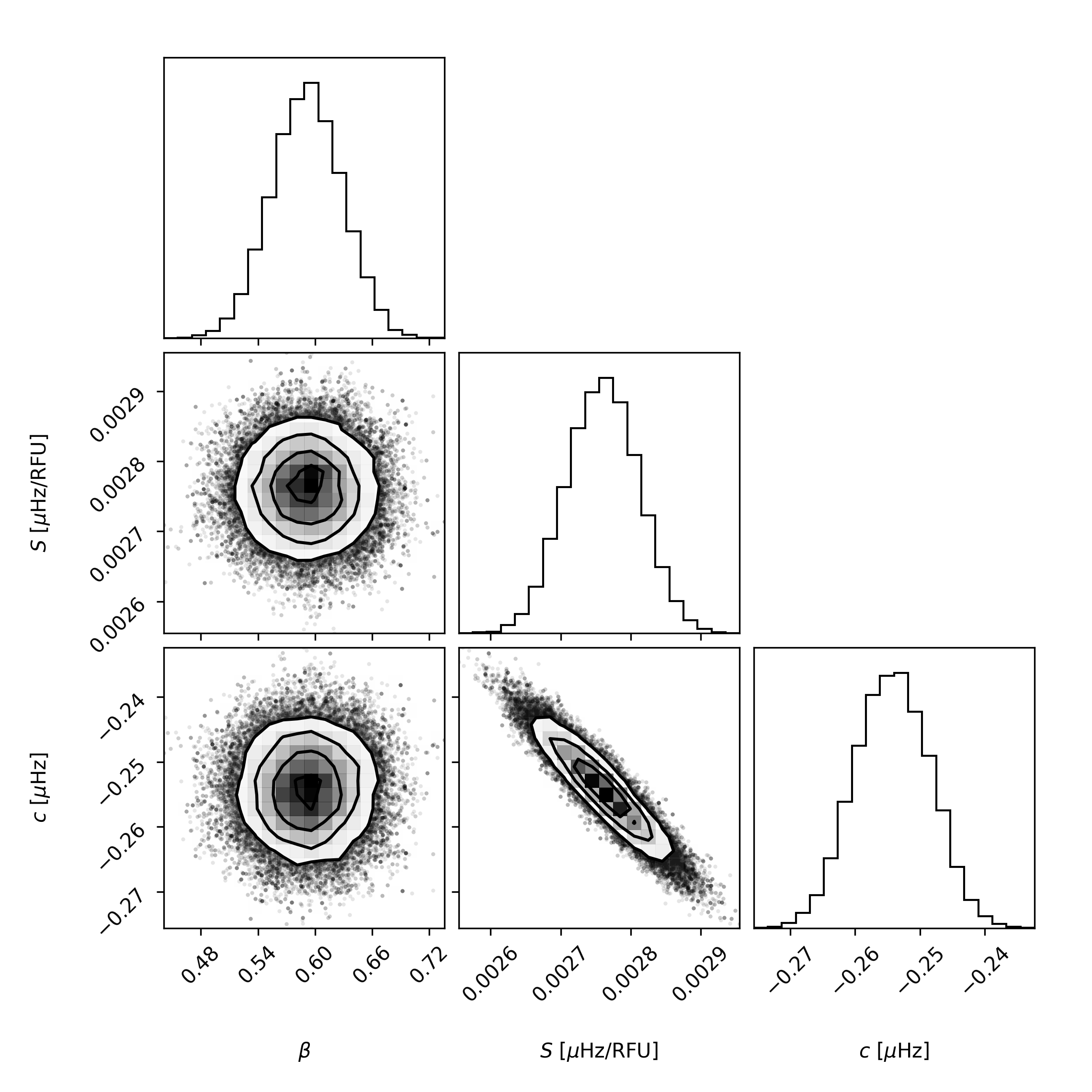}
\caption{Corner plot for fits to 7-d, non-overlapping frequency shifts using the two-sided activity index model of \ref{eq:twosides}.}
\label{fig:fig5}
\end{figure}

Figure~\ref{fig:fig5} shows the corner plot of posterior distributions from the fit. The fraction parameter $\beta$ is well constrained at $0.59\pm 0.04$ and not correlated with $S$ or $c$. This suggests that we can indeed improve the fit by including the effect of far-side activity.

\begin{figure}
\includegraphics[width=\linewidth]{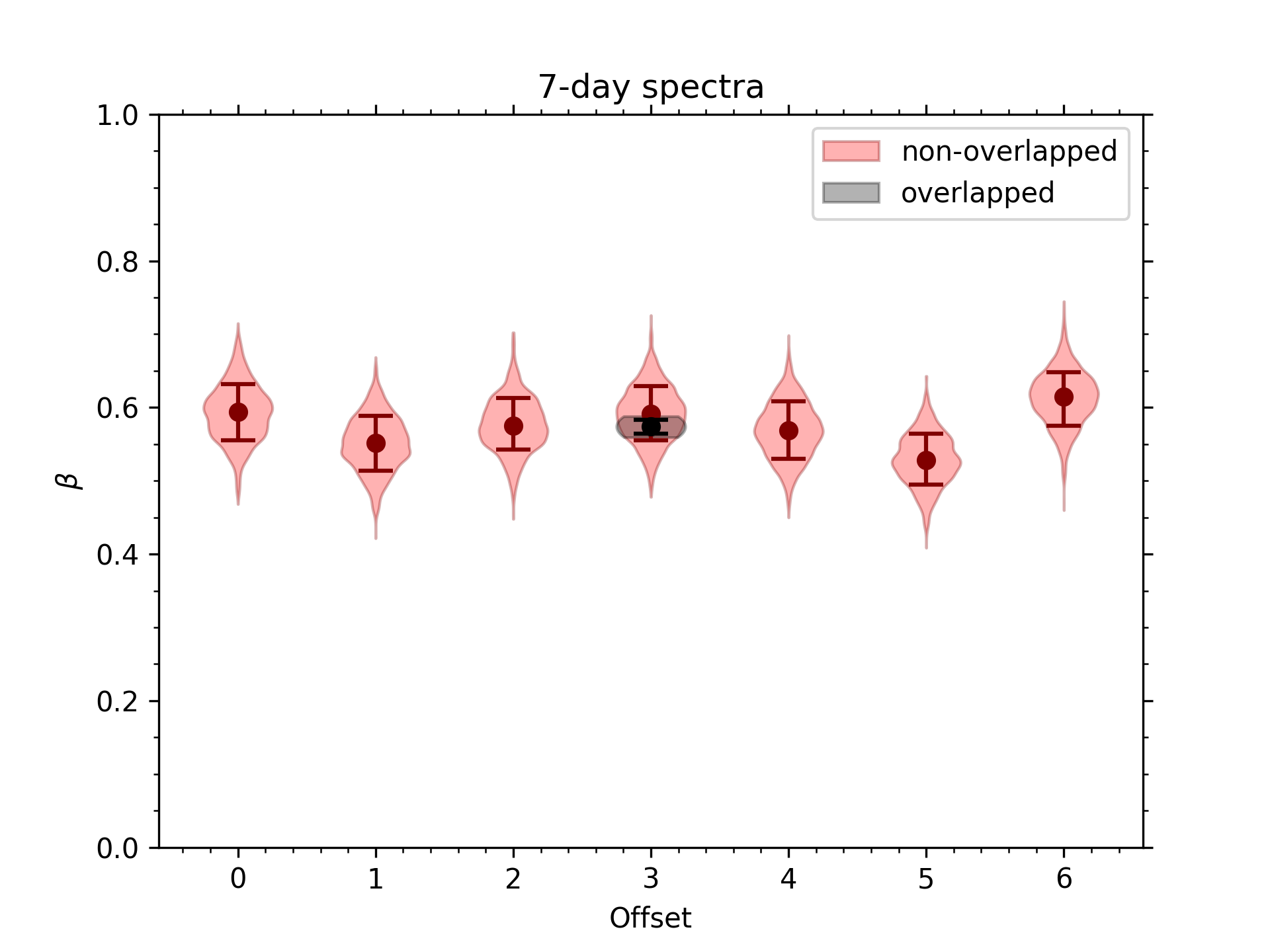}
\caption{Violin plots of the posterior distributions for $\beta$ for fits of Eq.~\ref{eq:twosides} to 7-d spectra with start Julian dates (JD) satisfying (JD - JD$_0$)  mod 7 = $i$ for $0\leq i \leq 6$, where JD$_0$ is the Julian date of the start of the overall time series. The distribution from fitting to the full set of overlapping spectra is shown in grey.}
\label{fig:fig7}
\end{figure}

To check that the apparent constraint on the value of $\beta$ is not a coincidence arising because of our choice of starting date for the non-overlapping subset of spectra, we repeat the fit to Eq.~\ref{eq:twosides} for spectra with start dates offset by 1 to 6\,d from the original choice. Figure~\ref{fig:fig7} shows the posterior distributions for $\beta$ for each offset value, together with the distribution obtained when fitting to the full set of overlapping spectra. The value obtained for $\beta$ is not significantly affected by the choice of offset.

\begin{figure}
\centering
\includegraphics[width=0.9\linewidth]{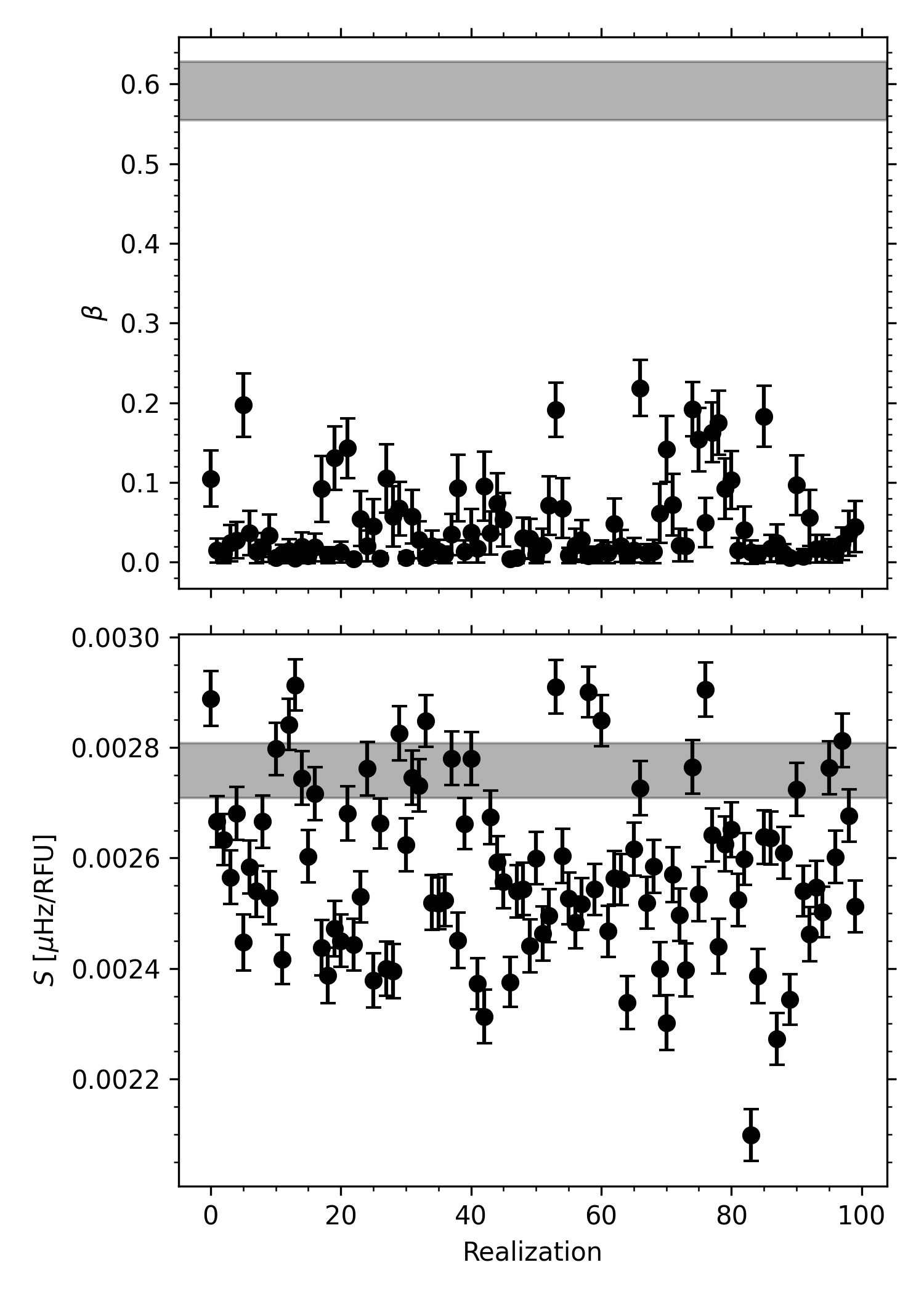}
\caption{Error-bar plots showing the summary statistics of the posterior distributions for the far-side fraction parameter $\beta$ and the sensitivity parameter $S$, for 100 realizations of synthetic data in which the residuals after a linear fit of non-overlapping 7-d frequency shifts to the near-side RF index $P_{\mathrm{near}}$ were randomly re-ordered and re-added to the near-side trend. The shaded grey area shows the  $\pm 1\,\sigma$ quantiles for the fit to the original data.}
\label{fig:fig6}
\end{figure}

In order to verify that what we interpret as the effect of including far-side activity is not simply a consequence of fitting the noisy data with a proxy that is smoother than the near-side RF index, we carried out a ``pre-whitening'' test as follows. First, the linear fit to the standard or near-side RF index (c.f. Eq.~\ref{eq:avshift}) was subtracted from the mean-frequency-shift data. The residuals were then randomly re-ordered and the near-side trend was added back to produce a synthetic data series, which was then fitted with the model of Eq.~\ref{eq:twosides}. This was repeated for 100 realizations. The distributions of the parameters $\beta$ and $S$ for each realization, together with those from the original data, are shown in Figure~\ref{fig:fig6}. The values for $\beta$ obtained from the reshuffled datasets are all close to zero, in contrast to the value from the original data, while for the sensitivity parameter $S$ the synthetic data gives values scattered around (but tending to lie slightly below) that for the original data. This gives us confidence that our $\beta$ value is a real feature of the data.

We have repeated the analysis with observing periods of different lengths. At lengths shorter than 7\,d, it is more difficult to obtain good frequency values because of the increasing coarseness of the frequency resolution. As the observation period increases towards 27\,d, the value obtained for $\beta$ becomes both more poorly constrained and more sensitive to the choice of start date. As can be seen from Figure~\ref{fig:proxy}, the front- and far-side proxies become less distinct as the observing period increases, so it is to be expected that our approach will work less well.

\section{Discussion}

One might expect that the influence of the far-side activity would be the same as that of the near-side activity, giving a $\beta$ value of 0.5 in Eq.~\ref{eq:twosides} instead of the $0.59\pm 0.04$ that we measure. 
One possible explanation is that Eq.~\ref{eq:twosides} underestimates the far-side activity level because it neglects activity that emerges and fades without rotating to face the Earth. While the effect of short-lived ephemeral regions on both the RF emission and the mode frequencies is believed to be small \citep[][]{2019MNRAS.489L..86C, 2013SpWea..11..394T}, our simple formula will also miss flux in strong regions that emerge on the far side of the Sun and significantly decay before rotating to face the Earth. A more sophisticated model would need to take into account the distribution of lifetimes and strengths of active regions; according to the review
by \citet{2015LRSP...12....1V}, small pores can have life-
times of days to weeks, while weaker ephemeral regions last hours
to days, but the lifetime roughly scales with the strength of the regions. The construction of such a model is beyond the scope of the current paper but is a potential avenue for future research.

\section{Conclusion and Prospects}

We have shown that it is possible to measure activity-related frequency shifts in BiSON observations from time series substantially shorter than one solar rotation. By constructing a simple proxy for far-side solar activity based on past and future observations of Earth-facing activity, we can estimate that the far-side activity appears to contribute slightly more than half of the overall change in the mode frequencies. Potentially, this means that BiSON frequency observations from short time-series could be used as an input to machine-learning models for predicting when new activity arises on the far side of the Sun, although the precision is not sufficient to see the effect of individual active regions.

While BiSON captures the lowest-degree modes, resolved helioseismic instruments such as GONG, MDI, and HMI can monitor global modes with degree up to a few hundred. Because each degree $l$ and radial order produces a multiplet of $2l+1$ rotationally split components, the central multiplet frequencies of higher-degree modes can be estimated with much more precision than is possible for BiSON's low-degree modes, potentially giving a much cleaner signal of the effect of far-side activity in spectra from short time series. \cite{2007SoPh..243..105T} examined the frequency shifts of GONG spectra from 9-d time series. They found that the frequencies in spectra for shorter time series were slightly less sensitive to the activity proxy than those from the usual 108-d GONG time series, which we suspect could be due to the overestimation of the activity changes when using a near-side-only activity proxy. In future work we will apply our analysis to such data. With the greater precision available from averaging many modes, it might be possible to see the effects of individual activity complexes. 

In principle it should be also be possible to perform an analysis of short datasets on solar-type stars that have high-quality asteroseismic observations. It would be necessary to obtain frequency estimates and activity proxies from spectra based on less than half a rotation period of observations. Targets observed photometrically that show pronounced signatures of rotational starspot modulation in their lightcurves would be of particular interest.

\section*{Acknowledgments}

We acknowledge the support of the UK Science and Technology Facilities Council (STFC) through
grant ST/V000500/1.

The peak-finding computations used in this paper were performed using the University of Birmingham's BlueBEAR HPC service, which provides a High Performance Computing service to the University's research community. See \url{http://www.birmingham.ac.uk/bear} for more details. 

This research made use of 
NASA's Astrophysics Data System.

We would like to thank all those who have been associated with BiSON over the years. 

We thank the anonymous referee for their helpful comments.

\section*{Software}
Below we include additional software used in this work which has not explicitly been mentioned above.
\begin{itemize}
    \item Python \citet{python1995}
    \item AstroPy \citep[][]{astropy:2013,astropy:2018}
    \item Matplotlib \citet{Hunter:2007}
    \item NumPy \citet{harris2020array}
    \item SciPy \citet{2020SciPy-NMeth}
    \item emcee \citet{2013PASP..125..306F}
    \item corner \cite{corner}
\end{itemize}

\section*{Data Availability}

 The BiSON time series analysed here is available at \url{http://bison.ph.bham.ac.uk/opendata}.

\bibliography{ms}
\end{document}